# Quasi-periodic changes of three dimensional solar anisotropy of galactic cosmic rays for 1965-2014

[1]Modzelewska R., [1,2]Alania M. V.
[1]Institute of Math. and Physics, Siedlce University, Siedlce, Poland
[2]Institute of Geophysics, Tbilisi State University, Tbilisi, Georgia
renatam@uph.edu.pl, alania@uph.edu.pl

\abstract
**Aims**
   {We study features of the three dimensional (3D) solar anisotropy of galactic cosmic rays (GCR) for 1965-2014 (almost 5 solar cycles 20-24). We analyze the 27-day variations of the two dimensional (2D) GCR anisotropy in the ecliptic plane, and the north-south anisotropy normal to the ecliptic plane. We study the dependence of the 27-day variation of the 3D GCR anisotropy on the solar cycle and solar magnetic cycle. We demonstrate that the 27-day variations of the GCR intensity and anisotropy can be used as an important tool to study solar wind, solar activity and heliosphere.}
**Methods**
   {We use the components $A_{r}$, $A_{\varphi}$ and $A_{t}$ of the 3D GCR anisotropy found based on hourly data of neutron monitors (NMs) and muon telescopes (MTs) using the harmonic analyses and spectrographic methods. We correct 2D diurnal ($\sim$ 24-hours) variation of the GCR intensity for the influence of the Earth magnetic field. We derive the north-south component of the GCR anisotropy based on the $GG$ index calculated as the difference in GCR intensities of Nagoya multidirectional MTs.}
**Results**
   {We show that behavior of the 27-variation of the 3D anisotropy verifies an existence of a stable long-lived active heliolongitudes on the sun. This finding illustrates usefulness of the 27-day variation of the GCR anisotropy as a unique proxy to study solar wind, solar activity and heliosphere. We distinguish a tendency of the 22-year changes of the amplitudes of the 27-day variation of the 2D anisotropy connected with the solar magnetic cycle. We demonstrate that the amplitudes of the 27-day variation of the north-south component of the anisotropy vary upon the 11 year solar cycle, however, a dependence of the solar magnetic polarity hardly can be recognized. We show that the 27-day recurrences of the $GG$ index and $A_{t}$ component are in a high positive correlation, and both are highly correlated with $B_{y}$ component of the heliospheric magnetic field.}

   \keywords{Sun: activity - Sun: heliosphere - Sun: solar wind - Sun: rotation

# 1 \section{Introduction}

Flux of galactic cosmic ray (GCR) measured at Earth consists of the isotropic and anisotropic parts. An isotropic part contains the various quasi-periodic changes with different time scales (from hours to several years), see e.g. (Kudela \& Sabbah \citeyear{Kudela16}, Chowdhury et al. \citeyear{CH16}, Bazilevskaya et al. \citeyear{Baz14} and references therein) and 3D spatial density gradients (Kozai et al. \citeyear{Kozai14}). An anisotropic part generally is reflected in the solar diurnal variation ($\sim$ 24 hours wave) assuming that while the Earth completes one rotation, a location of a source of anisotropic stream remains unchanged.
The mechanism of the solar diurnal anisotropy was explained by \cite{Ahluwalia62}, and a little later, by \cite{Krymsky64} and \cite{Parker64}, independently, based on the anisotropic diffusion-convection theory of GCR propagation in the heliosphere. \cite{Bieber93} have shed light on this problem assuming that 3D GCR anisotropy is a combination of the 2D solar ecliptic and the north-south anisotropies. 2D solar ecliptic anisotropy results in the daily variation of the count rate of ground based detector (e.g. neutron monitor (NM) or muon telescope (MT)) that rotates with Earth; the north-south anisotropy is revealing a flow of GCR normal to the ecliptic plane. \cite{Swinson69} proposed that the north-south anisotropy could have occurred due to the drift caused by positive heliocentric radial density gradient $\textbf{G}_{r}$ of cosmic rays and the $\textbf{B}_{y}$ component of the heliospheric magnetic field (HMF) (realized as the vector product, $\textbf{B}_{y}\times \textbf{G}_{r}$).

Among recent publications devoted to the GCR anisotropy one can mention, e.g., \cite{Kudela16}; \cite{Ahl15}, \cite{Munakata14}, \cite{Sabbah13} and \cite{Oh10}. However, quasi-periodic changes of the GCR anisotropy connected with the solar rotation, (further in this paper called the 27-day variation) were studied rarely until the work of \cite{Alania05, Alania08} for ecliptic plane anisotropy (2D case) and for the north-south component only for polar-located NMs by \cite{Owens80}. Swinson and coauthors (Swinson \& Yasue \citeyear{SY91, SY92}; Swinson at al. \citeyear{SYF93} and Swinson \& Fuji \citeyear{SF95}) reported the significant correlation between the 27-day variation of the north-south anisotropy for MTs data and the tilt angle of the heliospheric neutral sheet. \cite{Alania05, Alania08} and \cite{Gil12} analyzing the periods near the minima epochs of solar activity, demonstrated that the amplitude of the 27-day variation of the 2D anisotropy is less in the negative polarity period than in the positive polarity period of the HMF.

The main aim of this paper is two fold: (1) to investigate the 27-day variation of the 3D GCR solar anisotropy and its long-term changes during $\sim$ 5 solar cycles 20-24, and (2) to study the time lines of the 27-day variations of the north-south component of the GCR anisotropy and the 2D GCR anisotropy in the ecliptic plane.

**2 \section{Data and methods}**
We use data of neutron monitors: Kiel , Moscow, Oulu, Deep River and Climax with cut-off magnetic rigidities $R_{c}< 5$ GV to reveal explicitly an existence of the 3D GCR anisotropy. The pressure corrected hourly data of the GCR intensity were normalized as $I[\%]=\frac{x_{i}-\overline{x}}{\overline{x}}$, where $\overline{x}$ is average count rate of the observed GCR intensity. We excluded a trend larger than diurnal variation (24 hours) using the 25 hours moving average method.
As an example we present the hourly data of Moscow NM for January 2-11, 2007 in Fig. ~\ref{Fig1}, and the same 'detrended' data - in Fig. ~\ref{Fig2}.
Figures 1 and 2 demonstrate that the GCR intensity (Fig. ~\ref{Fig1}) alternates under an influence of trend with period larger than 25 hours, while the 'detrended' data (Fig. ~\ref{Fig2}) shows a fluctuation of the intensity around the zero level, which is essential to make a harmonic analysis available.

We calculate the daily radial $a_{r}$ and tangential $a_{\varphi}$ components of the diurnal variation of the GCR intensity by normalized and detrended hourly data of the GCR intensity using the harmonic analyses method (e.g. Gubbins

\citeyear{G04}):

$$I(t) = \frac{a_0}{2} + \sum_{k=1}^{\infty}(a_r^k \cos\frac{2\pi kt}{T} + a_\varphi^k \sin\frac{2\pi kt}{T}) = \sum_{i=1}^{\infty} a_k \sin(\frac{2\pi kt}{T} + \varphi_k) \quad (1)$$

where
$a_r^k = \frac{1}{p}\sum_{i=1}^{2p} x_i \cos\frac{\pi ki}{p}$, $a_\varphi^k = \frac{1}{p}\sum_{i=1}^{2p} x_i \sin\frac{\pi ki}{p}$, $a_0 = \frac{1}{p}\sum_{i=1}^{2p} x_i$,
$\varphi_k = arctg\frac{a_\varphi^k}{a_r^k}$, $a_k = \sqrt{a_r^2 + a_\varphi^2}$

$2p$=24 hours, $x_{i}$ designates the hourly data of the GCR intensity for each NM.
We correct the daily radial $a_{r}$ and tangential $a_{\varphi}$ components of the diurnal variation of the GCR intensity due to influence of the Earth magnetic field (Rao et al. \citeyear{Rao63}; Dorman et al. \citeyear{D72}; Dorman \citeyear{D09}) taking into account the asymptotic cone of acceptance characteristic for each NM. It is done by the rotation of the corresponding angle $\lambda$ (asymptotic longitude) for each NM. We calculate the radial $a_{r}^{E}$ and the tangential $a_{\varphi}^{E}$ components of the diurnal variation of the GCR intensity corrected for the influence of the Earth magnetic field using the expressions:

$$a_r^E = a_r \cos\lambda - a_\varphi \sin\lambda \qquad a_\varphi^E = a_r \sin\lambda + a_\varphi \cos\lambda \quad (2)$$

Furthermore, we exclude from consideration the amplitudes $>0.7\%$ as an anomalous events related to the disturbances in interplanetary space, generally, corresponding to the periods of Forbush decreases. A number of the excluded days is less than 2-3\% from the total number of days using for analyses.
The amplitude of the diurnal variation of the GCR intensity $a_{j}^{E}$ at any point of the observation (by NM) with the geomagnetic cut-off rigidity $R_{c}$ and the average atmospheric depth $h_{j}$ can be defined as:

$$a_j^E = \int_{R_c}^{R_{max}} \frac{\delta D(R) W(R, h_j)}{D(R)} dR \quad (3)$$

where $\frac{\delta D(R)}{D(R)}$ is the rigidity spectrum of the diurnal variation of the GCR intensity and $W(R,h_{j})=\frac{D(R)}{N}m(R,h_{j})$ is the coupling function (Dorman \citeyear{D63}; Yasue et al. \citeyear{Y82}); $R_{max}$ is the upper limiting rigidity beyond which the amplitudes of the diurnal variation of the GCR intensity vanish. For the power type of the rigidity spectrum $\frac{\delta D(R)}{D(R)}=AR^{-\gamma}$ one can write:

$$A_j = \frac{a_j^E}{\int_{R_c}^{R_{max}} R^{-\gamma} W(R, h_j) dR} \quad (4)$$

where $a_{j}^{E}$ is the observed amplitude of the diurnal variation of GCR intensity for $j^{th}$ NM, and $A_{j}$ - the corresponding amplitude of the anisotropy of GCR in the heliosphere. The values of the $A_{j}$ must be the same in the scope of the accuracy of the calculations for the arbitrary NM data when the pairs of the parameters $\gamma$ and

$R_{max}$ are properly determined. A similarity of the values of the $A_{j}$ found for different NMs is a decisive factor to affirm that the data of the given NM are trustworthy. We convert the radial $a_{r}$ and azimuthal $a_{\varphi}$ components of the diurnal variation of GCR into the radial $A_{r}$ and azimuthal $A_{\varphi}$ components of the GCR anisotropy in the heliosphere (free space) (Dorman \citeyear{D63}; Yasue et al. \citeyear{Y82}):

$$A_{r_j} = \frac{a_{r_j}^E}{\int_{R_c}^{R_{max}} R^{-\gamma} W(R, h_j) dR} \qquad (5)$$

$$A_{\varphi_j} = \frac{a_{\varphi_j}^E}{\int_{R_c}^{R_{max}} R^{-\gamma} W(R, h_j) dR} \qquad (6)$$

where the expression $\int_{R_{c}}^{R_{max}} R^{-\gamma}W(R,h_{j})dR$ is called the coupling coefficient (CC). This is a ratio of the observed amplitude of diurnal variation to the corresponding amplitude of the anisotropy of cosmic rays in the heliosphere. We provide a conversion for the same set of parameters as in \cite{Bieber91}, namely $R_{max}$ =100 GV and spectral index $\gamma$=0. This selection of the parameters $R_{max}$ and $\gamma$ is reasonable as far there is satisfied the criterion of the equality of the values of the $A_{j}$ (the amplitudes of the GCR anisotropy in the heliosphere) found for different NMs. We have found that for $\gamma$ = 0.5 there are not any valuable changes in results, i.e., that a rigidity dependence of anisotropy is reasonably weak for the energy range of GCR particles to which NMs respond; hard spectrum of GCR anisotropy was found in \cite{Hall96}.\\
Details of the NMs used in this study and values of corresponding $CC$ versus solar activity are presented in Table 1.

Table 1. Details of the neutron monitors and values of corresponding coupling coefficients (*CC*) vs. solar activity.

| NM station | latitude | longitude | $R_c$ [GV] | $\lambda$[°] | CC minimum | CC intermediate | CC maximum |
|---|---|---|---|---|---|---|---|
| Climax | 39.40N | 253.8E | 3.03 | -67 | 0.705 | 0.7105 | 0.716 |
| Deep River | 46.10N | -77.5W | 1.14 | -44 | 0.715 | 0.7215 | 0.728 |
| Kiel | 54.30N | 10.1W | 2.36 | 60 | 0.659 | 0.6355 | 0.612 |
| Moscow | 55.50N | 37.3E | 2.43 | 84 | 0.661 | 0.6380 | 0.615 |
| Oulu | 65.05N | 25.47E | 0.8 | 62 | 0.619 | 0.5895 | 0.560 |

We also use results of 3D GCR anisotropy calculated by the IZMIRAN group (Abunina et al. \citeyear{Abunina15}, Belov et al. \citeyear{Belov05}) applying the global spectrographic method (GSM) (Krymsky et al. \citeyear{Krymsky66, Krymsky67}) containing all operating NMs. Furthermore, to study the features of 3D GCR anisotropy in relatively wide range of GCR spectrum, we use data from Nagoya MTs (Munakata et al. \citeyear{Munakata14}) with median rigidity $\sim$ 60 GV.

In this paper we use the power spectrum density (PSD) method to reveal the quasi-periodicity in the analyzed time series; this method decomposes the time series into frequency ($\omega$ ) or period ($T$) components (e.g., Otnes and Enochson \citeyear{OE72}; Press et al. \citeyear{P02}):

$$\psi(\omega) = \int_{-\infty}^{\infty} R(t)e^{-i\omega t}dt \qquad (7)$$

where $R(t)$ is the autocorrelation function. In a discrete case we have

$$\psi(\omega) = \Delta t \sum_{r=-N}^{N} W \cdot R(r)e^{-i\omega r \Delta t} \qquad (8)$$

here $R(r)=\frac{1}{N-r}\sum_{i=1}^{N-r}x_{i}x_{i+r}$ is the autocorrelation function of time series $x_{i}$ and $W$ is the window function (we use Parzen's window function). The PSD of each frequency $\omega$ is calculated as $|\psi^{2}(\omega)|Hz^{-1}$.

**3 \section{27-day variation of the 2D GCR anisotropy in the ecliptic plane}**

Unfortunately, up to present it is not available a complete precise method for simultaneous calculation of a radial $A_{r}$, azimuthal $A_{\varphi}$ and latitudinal (north-south) $A_{t}$ components of 3D GCR anisotropy based on the world wide network of NMs and MTs. The $A_{r}$ and $A_{\varphi}$ components can be calculated using the both GSM and harmonic analyses methods based on data of NMs with cut-off rigidities $<5$ GV; in addition a latitudinal $A_{t}$ (north-south) component is possible to calculate in scope of the arbitrary constant, only by GSM method. Additionally, $A_{t}$ component can be estimated as a difference of two NMs located in regions of north and south poles (Chen and Bieber \citeyear{Bieber93}), also, $A_{t}$ can be estimated by directed MTs. However, these data are not homogenous and to study features of 3D anisotropy, (e.g. the 27 day variation of 3D anisotropy) we have to use results for $A_{r}$, $A_{\varphi}$ and $A_{t}$ obtained in different ways.

27-day variations of the GCR intensity (e.g., Richardson et al. \citeyear{Richardson99}, Dunzlaff et al. \citeyear{Dunzlaff08}, Guo \& Florinsky \citeyear{GF14, GF16}, Gil \& Mursula \citeyear{GM17}, Kopp et al. 2017) and anisotropy (e.g. Modzelewska \& Alania \citeyear{MA12}, Mavromichalaki et al. \citeyear{M16}) have a sporadic character. Their amplitudes significantly increase and decrease averagely during 4-6 solar rotation periods. However, they are not completely random phenomena. Generally, there always exists some levels of amplitudes of the 27-day variations of the GCR intensity and anisotropy being above the background fluctuations. This statement is demonstrated in Fig. ~\ref{Fig3}-\ref{Fig5}, where are presented changes of the GCR intensity (Fig. ~\ref{Fig3}), and $A_{r}$ (Fig. ~\ref{Fig4}) and $A_{\varphi}$ (Fig. ~\ref{Fig5}) components of the GCR anisotropy for the period of 2007-2009 for Oulu NM. We present in Fig. ~\ref{Fig3}-\ref{Fig5} corresponding to each period the same data 'detrended' by moving smoothing intervals of 29 days and filtered periodic oscillation with the band-pass period within 24-32 days. Figures ~\ref{Fig3}-\ref{Fig5} show that there exists periods with well pronounced 27-day recurrence with high amplitudes of the 27-day variations of the GCR intensity and the components of anisotropy, as well with very small amplitudes in other periods. This behavior of the 27-day variations of the GCR intensity and solar anisotropy is related with some disparate mechanisms. The 27-day variation of the GCR intensity in main is connected with heliolongitudinal asymmetry of solar wind and solar activity and their dependences on heliolatitudes, i.e. in creation of the 27-day variation of the GCR intensity contributes a fairly large part of the interplanetary space. At the same time, a convection – diffusion mechanism of solar anisotropy (Krymsky \citeyear{Krymsky64}, Parker \citeyear{Parker64}) is determined by the local processes near the Earth orbit and requires less part of the heliosphere. Above all, especially in the solar activity minima epochs there seldom is observed the 27-day variation of the anisotropy apparently connecting with drift of cosmic rays in the sector structure of the HMF, while there is observed a feeble the 27-day variation in the GCR intensity. There can be observed a vice-versa circumstances, when a clear 27-day variation of the GCR intensity is observed, not following with the valuable 27-day variation of anisotropy at all. Unfortunately, we have not completely explained this phenomena, which indicates that problem of the 27-day variations of anisotropy and intensity is not fully understandable. Up to now we have not got available an universal mechanism of the 27-day variations of the anisotropy and intensity of GCR account for dynamical changes of the features of the sources in the solar atmosphere and heliosphere. So, up to now there still remains much necessity to find new properties of the 27-day variations of anisotropy and intensity of GCR. We think that among others, this problem is an important aim of this paper.

Taking into account sporadic nature of the 27-day variation of GCR anisotropy (averagely lasting $\sim$ 4-6 solar rotations), there is of natural interest to study its behavior for a long period of 1971-2014. For this purpose we apply a spectral analysis method based on Oulu NM data. Results of calculations are presented in Figure ~\ref{Fig6}.
Figure ~\ref{Fig6} presents the spectral analysis of the daily radial $A_{r}$, azimuthal $A_{\varphi}$ components and phase$=arctan(A_{\varphi}/ A_{r})$ of the 2D GCR anisotropy for Oulu NM for long period 1971-2014. Figure ~\ref{Fig6} shows that $A_{r}$ and $A_{\varphi}$ components and phase demonstrate apparent quasi-periodic changes related to the Sun's rotation. For both components $A_{r}$ and $A_{\varphi}$, and phase the highest peaks correspond to the period of 25.6 days with 95\% confidence level. Fig. ~\ref{Fig6} demonstrates an existence of a stable long-lived active heliolongitudes on the sun which we consider as a source of the 27-day variation of the GCR anisotropy.

4 \section{27-day variation of the north-south anisotropy}

To study the 27-day variation of the north-south anisotropy we use two types of data: (1) daily $A_{t}$ component calculated by GSM for NMs data (http://cr0.izmiran.ru/CosmicRayAnisotropy/)
and (2) daily $GG$ index obtained for MTs (http://www.stelab.nagoya-u.ac.jp/).
$GG$ index (Mori \& Nagashima \citeyear {MN79}) is the difference between intensities recorded in the geographically north ($N2$) south ($S2$) and east ($E2$)-viewing directional channels for 49 degrees inclination corresponding to median rigidity $\sim$ 60 GV and calculated as follows: $GG=(N2-S2)+(N2-E2)$.

This method for studying the north-south asymmetry has an advantage of using GCR intensity data from a single location, instead of comparing GCR intensities from north and south polar NMs. The $GG$ index introduced according to the statement of (Mori \& Nagashima \citeyear {MN79}) should be free of noise in isotropic intensity caused by Forbush decreases, periodic variations, atmospheric temperature effects, and geomagnetic cut-offs. Although counting rate of GCR intensities in different directions ($N2, S2, E2$) could not precisely contain the same type of information, in spite $GG$ is accepted as a good alternative index by worldwide cosmic ray community (e.g., Munakata et al. \citeyear{Munakata14}).
Recently, we analyzed (Modzelewska \& Alania \citeyear{MA15}) a behavior of the quasi-periodic changes of the $GG$ index for 2007-2012 utilizing wavelet time-frequency method.

In this paper we extend this study of the $GG$ index and $A_{t}$ component of the anisotropy for 1971-2014 using PSD method.

We present the results of analysis in Figs. ~\ref{Fig7} and ~\ref{Fig8}. Figure ~\ref{Fig7} shows results of spectral analysis of $A_{t}$ component of the 3D GCR anisotropy for 1971-2006 (data of $A_{t}$ is available at the IZMIRAN website up to 2006) and Fig. ~\ref{Fig8} of $GG$ index for 1971-2014. Figures ~\ref{Fig7} and ~\ref{Fig8} show that both $A_{t}$ component and $GG$ index demonstrate apparent quasi-periodic changes related to the Sun's rotation, for $A_{t}$ component the period is 27 days and for $GG$ index 26.3 days with 95\% confidence level. **In Figs. 6 (left panel), 7 and 8 peaks at ~28-29 days are visible, but they are statistically insignificant-nevertheless, we don't exclude a presence of an additional source with an average period 28-29 days due to the differential rotation of the sun, manifested as a heliolongitudinal asymmetry versus heliolatitudes.**

5 \section{Long period changes of the amplitudes of the 27-day variations}
5.1 \subsection{2D ecliptic plane anisotropy}
To study long period changes of the amplitudes of the 27-day variation of the 2D ecliptic plane GCR anisotropy we calculate the amplitudes of the 2D 27-day variation of the anisotropy $(A27A)$ for Climax, Deep River, Kiel, Moscow, Oulu NMs data for 1965-2014. We use for calculations the formula (Alania et al. \citeyear{Alania05}) combining the 27-day recurrence of both radial and azimuthal components:

$$A27A = \sqrt{(A_{rr}(27) + A_{\varphi r}(27))^2 + (A_{r\varphi}(27) + A_{\varphi\varphi}(27))^2} \qquad (9)$$

here $A_{rr}(27)$ and $A_{r \varphi}(27)$ are the coefficients of the 27-day wave of the $A_{r}$ component, and $A_{\varphi r}(27)$ and $A_{\varphi \varphi}(27)$ are the coefficients of the 27-day wave of the $A_{\varphi}$ component.

Results of calculations we present in Fig. ~\ref{Fig9}. Figure ~\ref{Fig9} shows the temporal changes of the average amplitude of the 27-day variation of the 2D GCR anisotropy in the ecliptic plane smoothed over 13 solar rotations for Climax, Deep River, Kiel, Moscow, Oulu NMs in the time interval 1965-2014; error bars are calculated as standard deviations for considered NMs. Due to smoothing over 13 solar rotations the error bars are reduced, but at the same time, we lost information about quasi periodic changes for time interval less than one year.

Figure ~\ref{Fig9} demonstrates that the average amplitude of the 27-day variation of the GCR 2D anisotropy is larger in the minima periods of the positive ($A>0$) polarity epochs than in the negative ($A<0$) epochs, being in good agreement with \cite{Alania05, Alania08}. In addition the time line of the amplitudes of the 27-day variation of the 2D GCR anisotropy shows a tendency of the 22-year variation and a feeble, but systematic decreasing near the periods of solar minimum for $A>0$ polarity ($\sim$ 1975 and $\sim$ 1996), as well, indicating an existence of a weak 11-year solar cycle modulation.

The ecliptic 2D anisotropy components reflect significantly the drift pattern of cosmic rays flow and consequently the polarity dependence is apparent in the 27-day variation of the 2D GCR anisotropy.

To reveal a polarity dependence of the 27-day variation of the 2D GCR anisotropy we calculate the differences of amplitudes between $A>0$ and $A<0$ polarities which is equal to $\sim 40\%$ relative to the mean value. Particularly, average amplitude of the 27-day variation of the 2D GCR anisotropy $A27A$ for the whole period 1965-2014 is $(0.23\pm0.04)\%$, and values of $A27A$ for the consecutive minima are:

\\for $A>0$ 1975-1977 $A27A=(0.26\pm0.01)\%$
\\for $A<0$ 1985-1987 $A27A=(0.16\pm0.01)\%$
\\for $A>0$ 1995-1997 $A27A=(0.24\pm0.02)\%$
\\for $A<0$ 2007-2009 $A27A=(0.18\pm0.02)\%$

5.2 \subsection{Long period changes of the amplitudes of the 27-day variation of the north-south anisotropy}

Recently, we showed (Modzelewska \& Alania \citeyear{MA15}), that the daily $GG$ index is inversely related with the daily $B_{y}$ component of the HMF for 2007-2012. This effect, according to Swinson's (\citeyear{Swinson69}) formulation of the north-south anisotropy, is caused by drift, which we consider as an acceptable explanation. To study the nature of the north-south asymmetry we calculate the amplitudes of the 27-day variations of the $GG$ index ($A27GG$), $A_{t}$ ($A27A_{t}$) and $B_{y}$ components ($A27B_{y}$) using harmonic analysis method for 1971-2014. The results we present in Fig. ~\ref{Fig10}. Figure ~\ref{Fig10} demonstrates the temporal changes of the $A27GG$ and $A27A_{t}$ (top), and $A27B_{y}$ (bottom) for 1970-2014. A behavior of the amplitudes of the 27-day variations of the $GG$ index and $A_{t}$ component is similar and highly positively correlated with the 27-day variation of the $B_{y}$ component; all are alternating according to the 11-year solar cycle.

Contrary to the GCR anisotropy in the ecliptic plane, the north-south component has not got clear magnetic polarity dependence (e.g. Munakata et al. \citeyear{Munakata14}), so its 27-day recurrence also should not be dependent on the magnetic polarity, which we observe by experimental data.

6 \section{ Discussion and conclusions}

We study long term changes of the amplitudes of the 27-day variation of the 3D GCR anisotropy for 1965-2014. We show that a behavior of the 27-variation of 3D anisotropy verifies an existence of a stable long-lived active heliolongitudes on the sun. This find shows usefulness of the GCR anisotropy variation as an unique proxy to study solar wind, solar activity and heliosphere.

We recognize a tendency of the 22-year changes of the amplitudes of the 27-day variation of the 2D anisotropy connected with the solar magnetic cycle and a feeble 11-year solar cycle modulation.

We demonstrate that the amplitudes of the 27-day variation of the north-south component of the anisotropy vary in accordance with 11 year solar cycle, while there is hardly possible to show any dependence on the global solar magnetic field polarity. The 27-day recurrences of the $GG$ index and $A_{t}$ component are in a high positive correlation, and both are highly correlated with $B_{y}$ component of the HMF.

The observed properties of the radial $A_{r}$, azimuthal $A_{\varphi}$, normal $A_{t}$ components and $GG$ index demonstrate that quasi-periodic behaviors of the ecliptic 2D and north-south components of the 3D GCR anisotropy are governed by different mechanisms in the ecliptic plane and in the north-south direction. Due to complexity of the problem, we could not explain a full physical mechanism of the 27-day anisotropies in 3D space. We are not able to clarify completely processes related with the 27-day variations of 3D anisotropy using independently neither theoretical imaginations nor experimental data of NMs and MTs. For this purpose one has to consider together a theoretical modeling based on Parker's transport equation and experimental data.

\begin{acknowledgements}

We thank providers of data used in this study, especially OMNI database, IZMIRAN group and PIs of Oulu, Moscow, Kiel, Deep River, Climax NMs and Nagoya MTs. In composing a program for calculation of 2D ecliptic anisotropy components of GCR with the asymptotic cone of acceptance in Earth's magnetic field participated  Dr R. Aslamazishvili member of Prof. M. Alania's scientific group in Tbilisi. Remarks and suggestions of the Editor and Referee helped us to improve the paper.
\end{acknowledgements}

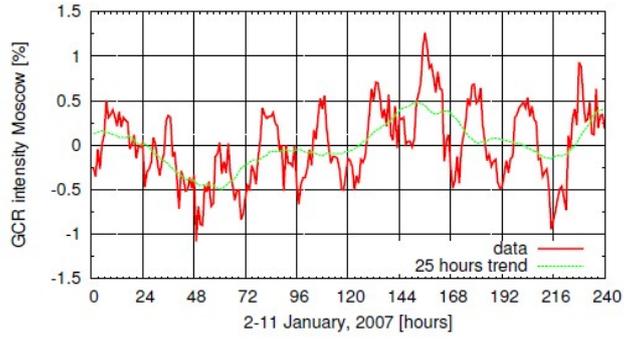

**Fig. 1.** Hourly data of Moscow NM in January 2-11, 2007 (red line) with the 25 hours trend (green line).

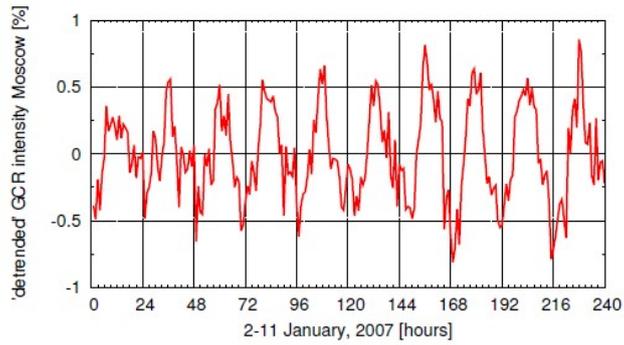

**Fig. 2.** The same data as in Fig 1 'detrended' with the 25 hours.

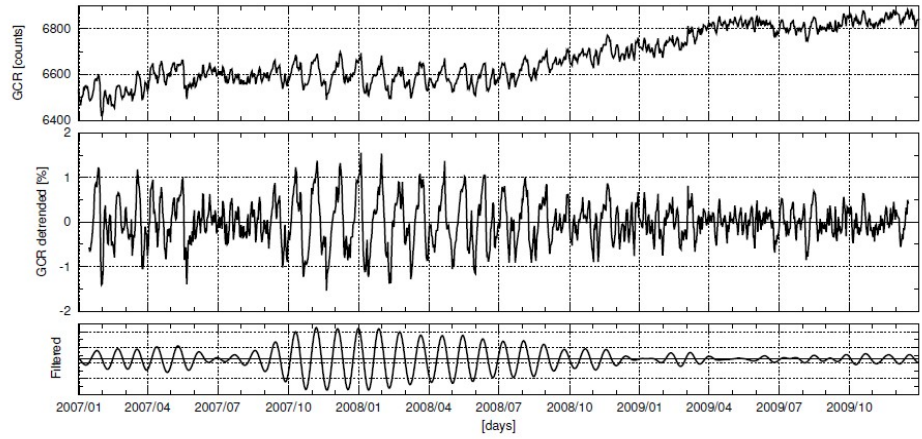

**Fig. 3.** Temporal changes of the daily GCR intensity for Oulu NM for the period of 2007-2009 (top) and corresponding to this period the same data detrended over 29 days (middle) and filtered periodic oscillation with the band-pass period within 24-32 days (bottom).

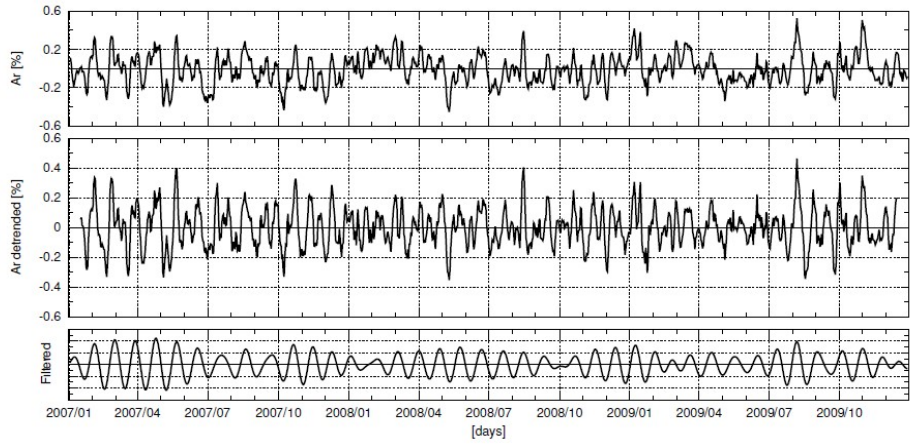

**Fig. 4.** The same as in Fig. 3 but for $A_r$ component of the GCR anisotropy for Oulu NM.

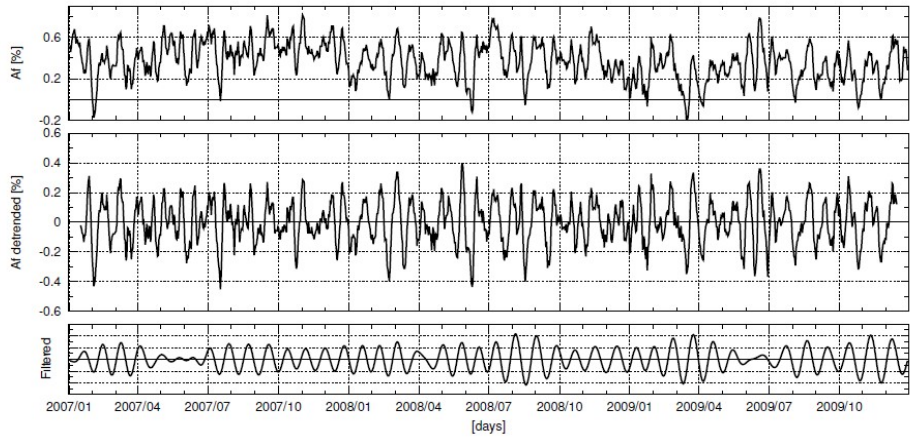

**Fig. 5.** The same as in Fig. 3 but for $A_\varphi$ component of the GCR anisotropy for Oulu NM.

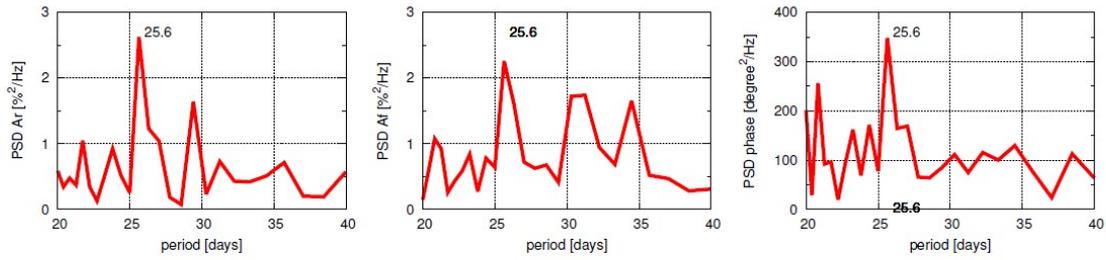

**Fig. 6.** Spectral analysis of the daily radial $A_r$ (left panel), azimuthal $A_\varphi$ (middle panel) and phase (right panel) of the 2D GCR anisotropy for Oulu NM for 1971-2014, for all the highest peak is for the period of 25.6 days with 95% confidence level.

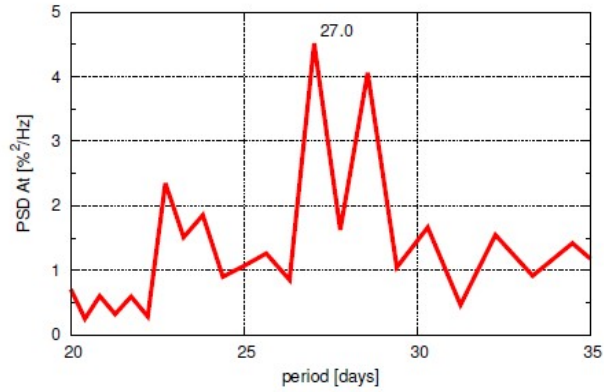

**Fig. 7.** Spectral analysis of the $A_t$ component of the 3D GCR anisotropy for 1970-2006, the highest peak is for the period of 27 days with 95% confidence level.

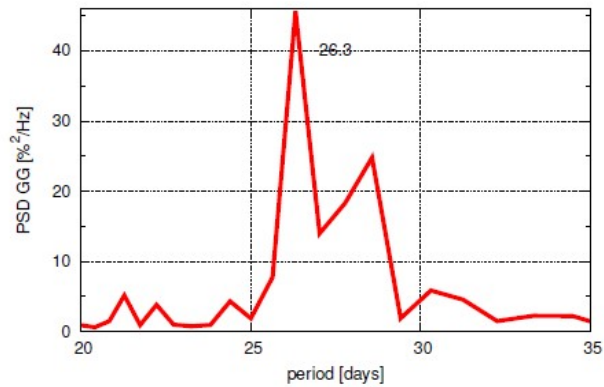

**Fig. 8.** Spectral analysis of the daily $GG$ index for 1971-2014, the highest peak is for the period of 26.3 days with 95% confidence level.

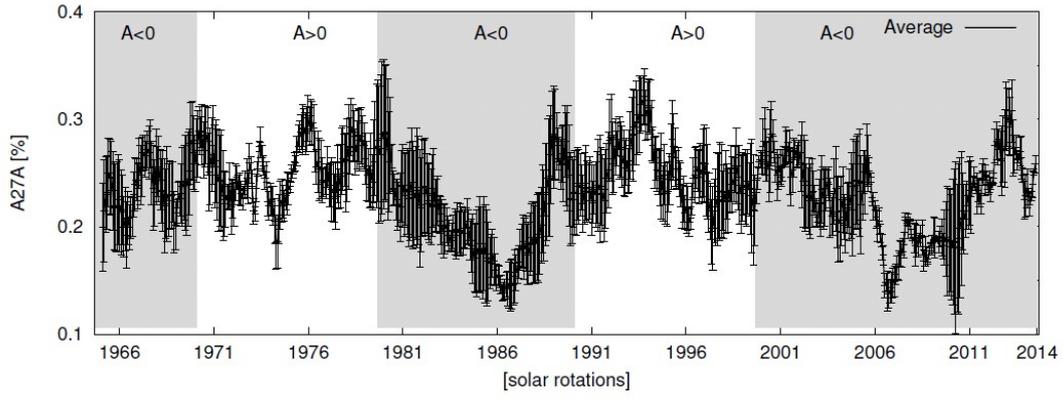

**Fig. 9.** Temporal changes of the average $A27A$ smoothed over 13 solar rotations for all considered NMs (Moscow, Kiel, Oulu, Deep River, Climax) for 1965-2014 during $A > 0$ and $A < 0$ polarity epochs; error bars are calculated as standard deviations for considered NMs.

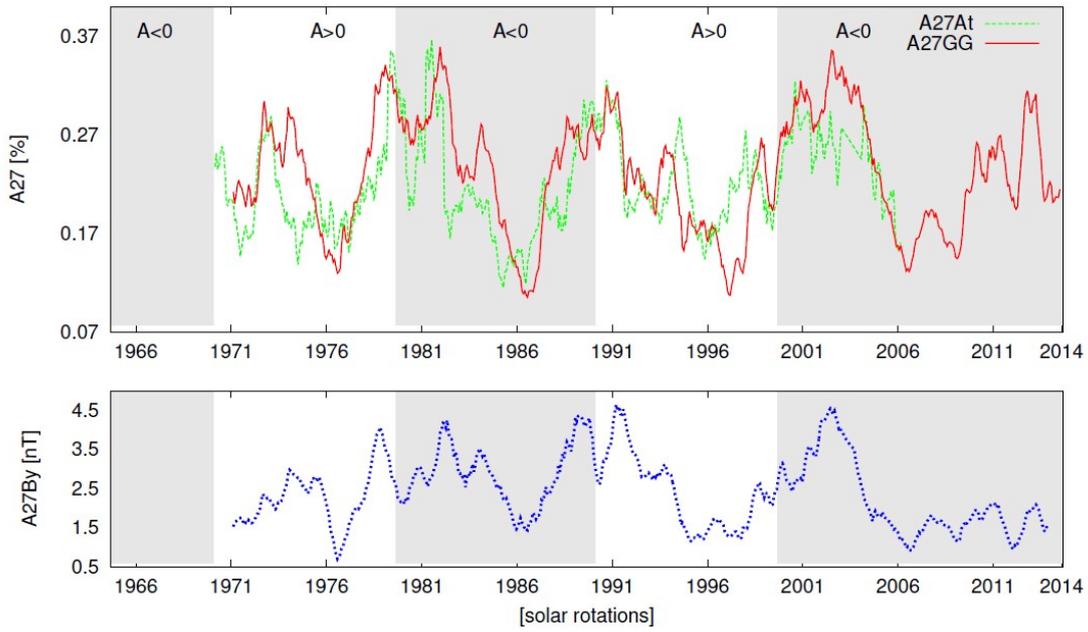

**Fig. 10.** Temporal changes of the amplitudes of the 27-day variation of the $GG$ index ($A27GG$) and $A_t$ component ($A27A_t$) of the 3D anisotropy (top) and $By$ component of the HMF ($A27B_y$) (bottom) smoothed over 13 Sun's rotations during $A > 0$ and $A < 0$ polarity epochs. Values $A27GG$ and $A27B_y$ are presented for 1971-2014, but $A27A_t$ – for 1970-2006, data of $A_t$ is available at the IZMIRAN website up to 2006.